\documentclass{article}
\usepackage{graphicx}
\usepackage[legalpaper, margin=1.25in]{geometry}
\usepackage{authblk}

\usepackage{lineno}

\usepackage{amsfonts}
\usepackage{amsmath}
\usepackage{amssymb}
\usepackage{amsthm}
\usepackage{algorithmic}
\usepackage{algorithm}
\usepackage{bm}
\usepackage{quantikz}
\usepackage{mathastext}
\usepackage{mathtools}
\usepackage{physics}
\usepackage{bbm}
\usepackage[sortcites,sorting=none,backend=biber]{biblatex}
\usepackage{multicol}
\usepackage{placeins}
\usepackage{caption}
\usepackage{subcaption}
\usepackage{dsfont} 
\usepackage{hyperref}
\usepackage[capitalize,noabbrev]{cleveref}

\newcommand{\E}{\mathbb{E}}
\newcommand{\Hc}{\mathcal{H}}
\newcommand{\Oc}{\mathcal{O}}
\newcommand{\Pc}{\mathcal{P}}
\newcommand{\KL}{D_{\text{KL}}}
\newcommand{\Id}{\mathds{1}}

\newtheorem{defn}{Definition}

\begin{document}

\tikzset{every picture/.style={line width=0.75pt}} 

\title{Approximating the Mathematical Structure of Psychodynamics}
%
%
\small
\author[1,2,3]{Bryce Allen Bagley}
\author[1,2]{Navin Khoshnan}

%
%
\affil[1]{Mathematical Medicine Group, Department of Neurosurgery, Stanford University, Stanford, CA, USA}
\affil[2]{Petritsch Lab, Department of Neurosurgery, Stanford University}
\affil[3]{Physician-Scientist Training Program, Stanford University}
%
\maketitle              
\begin{abstract}

The complexity of human cognition has meant that psychology makes more use of theory and conceptual models than perhaps any other biomedical field. To enable precise quantitative study of the full breadth of phenomena in psychological and psychiatric medicine as well as cognitive aspects of AI safety, there is a need for a mathematical formulation which is both mathematically precise and equally accessible to experts from numerous fields. In this paper we formalize human psychodynamics via the diagrammatic framework of process theory, describe its key properties, and explain the links between a diagrammatic representation and central concepts in analysis of cognitive processes in contexts such as psychotherapy, neurotechnology, AI alignment, AI agent representation of individuals in autonomous negotiations, developing human-like AI systems, and other aspects of AI safety.

\end{abstract}

\section{Introduction}

Across the breadth of psychological and psychiatric conditions where therapy has been shown to be an effective intervention, modalities within the category of psychodynamic psychotherapy (PDP) have proven broadly applicable and effective. However, a key factor adding difficulty to research on PDP is the difficulty of quantifying what PDP is \textit{doing} in a process sense. Forms of therapy such as cognitive behavioral therapy (CBT) can have discrete, comparatively easily quantifiable units, such a number of exposures to a distressing stimulus, but PDP in some sense involves a great deal more information. A range of psychometrics have been applied to successful studies on PDP, and the insights produced are of great value. Yet endpoint and outcomes data are limited in that they describe a snapshot of mental states. The ability to quantify the dynamics underlying PDP could open the door to a range of new directions in research on optimizing and modifying PDP. 

Many of the notions from PDP have the potential for substantial relevance to research questions related to human cognition in other, emerging fields. Cognition represents a critical attack surface for AI-powered attacks on neurotechnology,\cite{Denning_2009_neurosecurity_first_paper_neurosurgery_journal,Roelfsema_2018_mind_reading_and_writing_future_of_neurotech,bernal_2020_brain_implant_neuronal_cyberattacks_experiments_1,bernal_2022_bci_security_review,Bernal_2023_reasons_to_prioritize_BCI_security}  and also one of the attack surfaces which current methods are least equipped to protect.\cite{bagley_petritsch_2024_holographic_cognition_original_cognitive_neurosecurity} Likewise, in other realms of AI safety there is concern for the abilities of LLMs and other models to manipulate human cognition,\cite{zhang_2024_llm_manipulating_users,choi2024llme_manipulating_humans_or_humans_using_LLMs,sun2025friendlyfriendsllmsycophancy_manipulate_users,williams2025targetedmanipulationdeceptionoptimizing} and analysis and quantification of static states represents only part of the picture of human cognition. Recent work in this second area has provided timely insights into risks associated with human cognition as an attack surface for AI,\cite{zhang_2024_llm_manipulating_users,choi2024llme_manipulating_humans_or_humans_using_LLMs,sun2025friendlyfriendsllmsycophancy_manipulate_users,williams2025targetedmanipulationdeceptionoptimizing} but been largely heuristic in nature, and when modeling human cognition quantitatively relied on classical probability, which is well-established to yield incorrect results.

In recent decades an extensive body of literature in cognitive science and mathematical psychology has demonstrated a superficially particularly odd feature of human cognition. When one attempts to explain many features of cognition using classical probability, paradoxical outcomes are found in experimental studies. Yet by switching from classical probability to quantum probability, these paradoxes resolve in an elegant and empirically validated manner.\cite{Pothos_2009_violations_of_quote_rational_unquote_decisions_quantum_cognition,Trueblood_2012_quantum_cognition_causal_reasoning,Wang_2013_quantum_question_order_model,Wang_2013_the_potential_of_quantum_cognition,Wang_2014_context_effects_quantum_cognition,Bruza_2015_quantum_cognition_review,Busemeyer_2015_quantum_cognition_applied_to_psychology,Yearsley_2016_quantum_cognition_decision_theories,Pothos_2022_quantum_cognition_review,Busemeyer_2023_quantum_cognition_models,} It cannot be overly emphasized that quantum cognition has absolutely no connection whatsoever with unsupported hypotheses or pseudoscientific proposals asserting consciousness to be a quantum phenomenon. The sole link between quantum cognition and quantum mechanics is purely mathematical, in that they both use an alternative form of probability and statistics which is based on complex projective geometry (a different area of mathematics than the standard form of probability).\cite{von_neumann_1955_quantum_probability}

In a recent paper, Bagley \& Petritsch described how to take existing modeling approaches from the quantum cognition literature and extend it to the more general, high-dimensional case.\cite{bagley_petritsch_2024_holographic_cognition_original_cognitive_neurosecurity} They termed this \textit{holographic cognition} on account of its connections with hyperdimensional computing (centered on what are termed \textit{holographic representations} in the hyperdimensional computing literature), area of computing that also centers on high-dimensional data.\cite{klyeko_survey_hyperdimensional_computing_part_I,klyeko_survey_hyperdimensional_computing_part_II} The structure of holographic cognition is very well-suited to offering a means of quantifying PDP--and psychodynamics more broadly--at a level of complexity able to capture rich psychological features. However, the presentation of holographic cognition in that paper is in a form typical for papers on mathematics and its applications. We recognized that this approach would greatly impair collaboration between researchers who would take a more mathematical approach and those from less quantitative or non-quantitative backgrounds, in addition to essentially crippling much of its potential for broader utility.

What is needed is a new way of describing concepts in holographic cognition and standard quantum cognition. This means of description requires two core properties. First, it must be capable of expressing mathematical statements in a precise and rigorous manner, so that those studying the approach from a quantitative perspective can guarantee their results are likewise precise and rigorous. Second, it must be presented in such a way that it is \textit{equally} approachable for experts from many fields--and ideally even reasonably approachable for laypeople. If achieved effectively, the combination of these two features would provide a means for scientists, mathematicians, policy experts, and more to apply ideas and results from holographic cognition while all "speaking the same language", in some sense. 

To accomplish this task we make use of a special case of the mathematical field of category theory.\cite{coeke_kissinger_2017_picturing_quantum_processes_textbook} Specifically process theory, an approach to the study of systems via the framework of a Symmetric Monoidal Category (SMC). There already exists a reformulation of quantum mechanics in terms of process theory, which conveniently provides some of the mathematical apparatus required for our objective.\cite{coeke_2003_process_technical_report_logic_of_entanglement,abramsky_coeke_2004_process_categorical_semantics_quantum_computing,coeke_2005_process_kindergarten_quantum_mechanics,abramsky_coeke_2005_process_applications_of_categories_abstract_physics,baez_2006_process_quantum_quandries_categorical_perspective,selinger_2007_process_dagger_compact_closed_categories,baez_lauda_2011_process_quantum_world_mathematical_innovation,baez_stay_2011_process_physics,heunen_et_al_2021_process_quantum_physics_and_linguistics} A great many mathematicians and physicists have contributed to the development of that process theory,\cite{penrose_1971_process_original_paper_on_tensor_diagrams,bohm_peat_1987_process_implicit_order_time_space,abramsky_coeke_2004_process_categorical_semantics_quantum_computing,coeke_2005_process_kindergarten_quantum_mechanics,abramsky_coeke_2005_process_applications_of_categories_abstract_physics,baez_2006_process_quantum_quandries_categorical_perspective,selinger_2007_process_dagger_compact_closed_categories,baez_lauda_2011_process_quantum_world_mathematical_innovation,baez_stay_2011_process_physics,heunen_et_al_2021_process_quantum_physics_and_linguistics}  and to demonstrating that it is mathematically rigorous.\cite{joyal_street_1991_process_rigorous_quantum_tensor_calculus_geometry,joyal_street_verity_1996_process_rigorous_traced_monoidal_categories,hasewaga_et_al_2008_process_rigorous_finite_dimension_vector_symmetric_monoidal_categories,selinger_2011_process_rigorous_graphical_languages_monoid_categories,kissinger_2012_process_rigorous_pictures_of_processes_thesis} The results of their efforts are invaluable here. However, while their work provides those elements of a process theory required for the use of quantum probability, it is necessary to construct a new process theory describing psychodynamics due to key differences between cognition and quantum mechanics. While both operate based on complex projective probability, many phenomena are not shared. Development of this psychodynamics process theory is the central task of this paper. 

\section{Methods}
\subsection{Necessary Properties of a Process theory}\label{section_methods_key_properties_process_theory}

Process theories shift the focus of mathematical models away from the states occupied by systems at varying times, instead emphasizing the processes which act to evolve states over time. In this way, they are quite naturally suited to how one reasons clinically about human psychology and psychopathology. There are three features required of any process theory, as summarized by Coeke and Kissinger.\cite{coeke_kissinger_2017_picturing_quantum_processes_textbook} In summary, processes act on types, and one can link together networks of processes so long as the correct types serve as inputs and outputs to those processes.

\begin{enumerate}
    \item A collection $T$ of \textit{types} represented by wires.
    \item A collection of $P$ of \textit{processes} represented by boxes, where for each process the input and output types are from the collection $T$.
    \item A means of linking processes, which is an operation that interprets a diagram of wires and processes in the collection $P$ such that the entire system represents a process in $P$. 
\end{enumerate}

A crucial consequence of this is that for a diagram comprised of collections of processes, that diagram as a whole will always be a process itself. In mathematical terms, it is closed under compositions of processes. In clinical terms, a sequence or grouping of psychodynamic and therapeutic processes for a given individual are collectively a psychodynamic process for that same individual.  

\subsection{Constructing a Process Theory Description of Psychodynamics}\label{section_methods_constructing_PDPT}
Concepts from quantum cognition translate quite directly to the central ideas of psychodynamics and psychotherapy. \textit{Cogits}, proposed by Bagley and Petritsch as a term for the elementary units of a mental state, serve as a shorthand for the variety of things which can be represented in quantum cognition. Cogits can represent any of preferences, beliefs, memories, subconscious components of a mental state, and more. Complex instances of such facets of cognition can be represented by collections of cogits, for example by using some number $n$ cogits to describe a more nuanced or complex belief. Because all can be described with the same mathematics, a common name frequently helps clean up discussion of the ideas. Taking these ideas and the requirements for a process theory as delineated by Coeke and Kissinger,\cite{coeke_kissinger_2017_picturing_quantum_processes_textbook} we develop what follows.

There are at least two ways one could define the elements of the set $T$. Perhaps the most immediately intuitive option is to have $T$ include vectors of cogits,\footnote{For clinical or other non-quantitative readers: A vector is an ordered list of values. For example, if one listed out their favorites in each of various categories of foods, you would describe this information using a vector of some number of cogits.} vectors of stimuli coming from the outside world, and vectors describing the various outputs of cognition such as speech, movement, other actions, and so on. However, we argue that when conceptualized as processes which alter a person's mental state, the symmetry between internal dynamics of cognition and dynamic changes when exposed to external stimuli is such that stimuli are better represented as processes in the set $P$ within a psychodynamics process theory. 

As such, we instead use the following elements for a psychodynamics process theory. 

\begin{itemize}
    \item The types in $T$ are:
    \begin{itemize}
        \item Cogit state vectors representing relevant information about an individual's mental state at some time $t$. One need not describe the entirety of a person's cognition, and indeed this would be completely infeasible, so the size and content of these vectors is limited to what is sufficiently relevant to a given case. In cases where one needs to distinguish between cogit vectors, these vectors will be denoted by $c$ with subscripts.\footnote{For clinical readers: When there is a set of the same sort of mathematical object, the convention is to distinguish them with subscripts like $x_1$, $x_2$, ..., $x_{19}$, and so on. When you wish to distinguish between multiple items without specifying an exact item of that type, you would instead use a variable for the subscript, such as $C_i$, $C_j$, $C_k$ to denote three separate but yet-unspecified cogit state vectors.}
        \item Outputs of cognition, such as speech, actions, and so on. These will be denoted as $a$ with subscripts in the same fashion as cogit vectors.
    \end{itemize}
    \item The processes in $P$ fall into a few key categories:
    \begin{itemize}
        \item Internal measurement processes, where one is consciously aware of their own thoughts--i.e. taking internal measurements of some limited portion of their cogit state vector--and the changes occurring in their mental state. These will be denoted by $K$, sometimes with an additional subscript.
        \item External measurement processes not related to communication from another entity, where one's cognition is altered by some external stimuli in a fashion which may or may not generate a response $a$. These processes will be denoted by $E$, with or without subscripts.
        \item External measurement processes related to communication from another entity, which prompt a response. We denote these as $O$. 
        \item Cognitive dynamics separate from $K$. These will be denoted by $S$, with or without subscripts.

    \end{itemize}
\end{itemize}

In many cases there will be overlap between these types of processes or ambiguity in how to represent them, so the use of notation is flexible as we will see in the below diagrams. 

We now turn to showing the graphical notation for these same ideas. Let us begin with perhaps the simplest case, with the evolution of a mental state over time via subconscious processes. The diagram on the left below shows a cogit state vector $c$ evolving over time via subconscious processes $S_t$ at times $t=1, 2, 3$.

The scale of time over which these processes take place can be chosen on the basis of the use case. In tests of instinctive reactions, they might be separated by seconds. In the context of psychotherapy looking at how a person's behavior, nonverbal communication, and such progress over time, they could instead represent time spans of weeks or even months. Of note, the process \textit{itself} occurs over some time span. So $S_1$ could represent the impact of subconscious processes over seconds, weeks, months, or whatever time scale is most relevant. This idea is perhaps clarified by representing stimuli as an external measurement process, as in the middle diagram.

\FloatBarrier

\begin{figure}[h!]
    \centering
\begin{subfigure}[c]{0.25\textwidth}
    \includegraphics[width=0.3\textwidth]{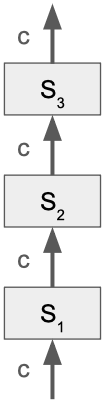}

\end{subfigure}
\begin{subfigure}[c]{0.25\textwidth}
    \includegraphics[width=0.49\linewidth]{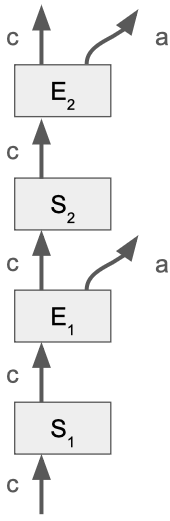}
\end{subfigure}
\begin{subfigure}[c]{0.25\textwidth}
    \includegraphics[width=0.56\linewidth]{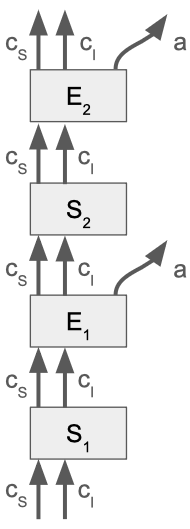}
\end{subfigure}

\end{figure}

\FloatBarrier

One has the option to use subscripts for further clarification, such as $c_0$, $c_1$, $c_2$, and $c_3$ to denote the initial state and those resulting from each of the subconscious processes. Likewise, $a_1$, $a_2$, and $a_3$ for the outputs from the external measurement operators. Concretely, the patient's words and behavior in response to three stimuli. These stimuli could represent part of an experiment in cognitive science, the effects of entire therapy sessions, or even a sequence of individual statements by a therapist during a single session. As previously mentioned, the timescale can be adjusted according to whatever is more useful for a given application. 

Suppose we wanted to distinguish between the portion of an individuals cogits representing conscious aspects of their mind and the portion representing unconscious aspects. We would then use parallel wires, and we selected $c_S$ and $c_I$ to respectively denote the subconscious portion of their mental state and the portion of which a person has conscious internal awareness. This can be seen in the diagram on the right above. However it is not so simple to delineate between the conscious and subconscious portions of a mental state, as the very same information could have a conscious effect on actions and thoughts in one setting and a subconscious impact in another. Because of this, we do not separate the two except for the purpose of illustrating the above idea. In the formulation of quantum probability via a SMC, this separation would require the conscious and subconscious cogits be \textit{bipartite}, meaning that there is no entanglement between the two subsets. It is a basic fact of psychology that this is incorrect, which is why we do not distinguish between the two and can in fact abandon the use of the $c$ label for cogit vectors in general. Likewise, we will later see that we can equally drop the $a$ notation for outputs. 

One could add subscripts indicating time-points for each of the $c$ and $a$ wires, but this is not needed. One of the helpful facets of these diagrammatic representations is that the distinction between different $c$ and $a$ (as well as different $S$, $E$, and $I$) is made clear through the diagram itself. By having the lower and upper $a$ be separate wires, we are inherently stating them to be different things, albeit of the same type. Likewise, we could remove the subscripts from the $S$ and $E$ processes seen here without losing any of what the diagram conveys. 

However, it should be noted that the mathematical properties of these diagrams are such that the horizontal and/or vertical ordering of wires' entries and exits from processes are not interchangeable. This property is displayed in the examples below.

\begin{figure}[h!]
    \centering
    \includegraphics[width=0.7\textwidth]{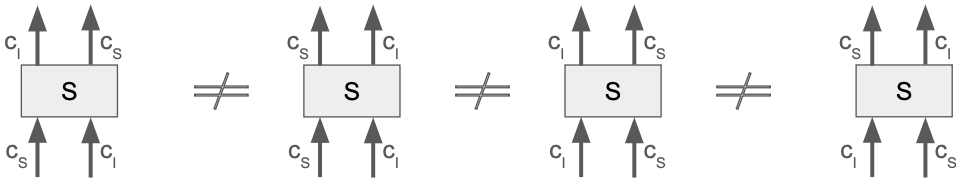}
\end{figure}
\FloatBarrier

While thus far we have used arrows to indicate the direction of time, we have the option of removing them and simply using the standard formation of time progressing upwards along a diagram. For the sake of clarity, however, we will continue to use arrows for the time being. 

Finally, we describe the diagram features used to denote probability. In the process theory for quantum probability, it is standard to use triangles to denote states, with downward-pointing triangles reflecting the cogit vector(s) state(s) at the start of some process being considered, and upward-facing triangles reflecting probabilities of measuring various options from the set of possible states. 

For example, suppose a case where we simply wish to describe the probability of measuring a cogit vectors as being in some specific state, absent the influence of whatever means of measurement was used as a readout. This is of course not possible in reality, and is shown here for the sake of simplicity.

\begin{figure}[H]
    \centering
    \includegraphics[width=0.08\linewidth]{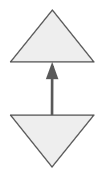}
\end{figure}

\subsection{Multiparty Processes}

We now turn to the case which is more interesting for our purposes: that of diagrams representing interactions between individuals and/or entities of other sorts. 

Take, for example, the following diagram:

\begin{figure}[H]
    \centering
    \includegraphics[width=0.5\linewidth]{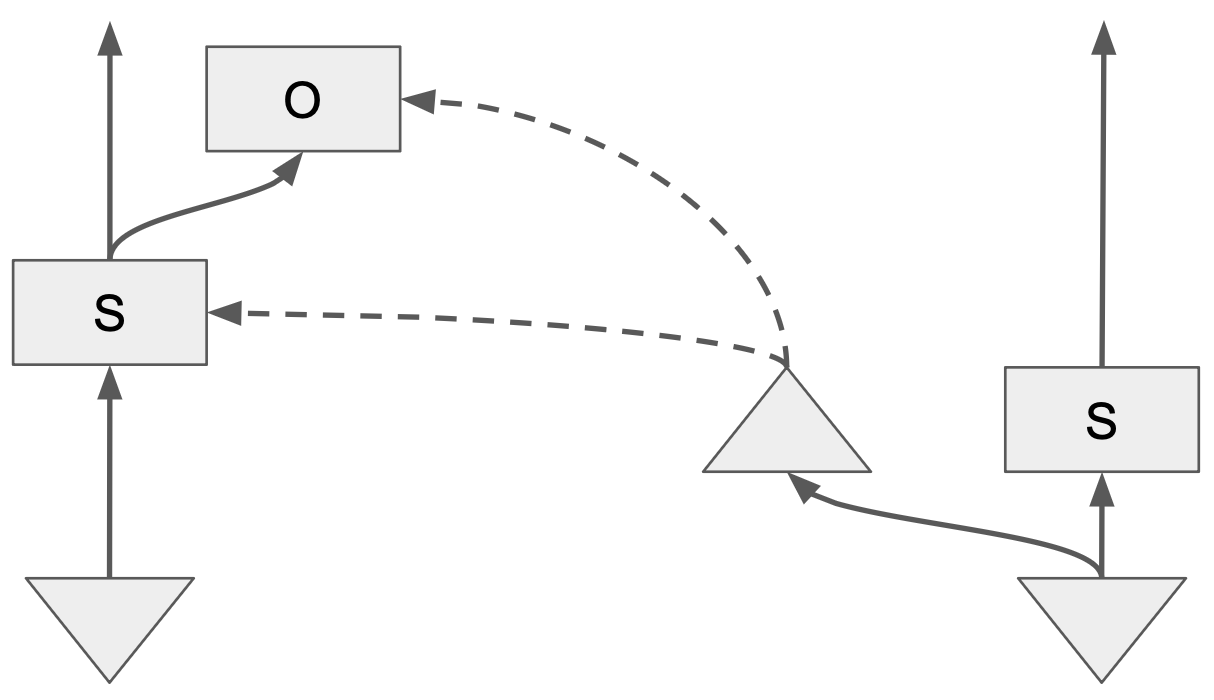}
\end{figure}

At the start, each party's initial state is respectively represented by the pair inverted triangles. Both have some generic processes $S$ by which their cogit vectors evolve, and for all $S$ in this diagram we denote generic processes rather than specific and identical ones. Some output determined probabilistically by the initial state of the right-side party (RSP) initial state produces an effect on the left-side party (LSP). While the right-side party's cogit vector continues to evolve on its own, the probabilistic output of their initial state then applies two operators to the LSP's cogit vector. First, some operator by which the output from the RSI influences the evolution of the LSP's cogit vector. Second, some operator $O$ which reflects a measurement. The dotted lines reflect the causal links between the measured output values--or more generally the distribution over possible measured outputs--and the $\{S,O\}$ pair which any given output applies to the other party's cogit vector. The fact the outputs reflect a distribution over possible cogit vector states means that the paired $S$ and $O$ induced by an output are themselves operator-valued distributions (OVDs), which will be discussed at greater length later. 

In the following diagram, we will see a reciprocal process induced by these output $\{S,O\}$ pairs. To help visually distinguish between the evolution of states over time vs the application of operators due to outputs, we use dashed lines to link the latter. 

\begin{figure}[H]
    \centering
    \includegraphics[width=0.5\linewidth]{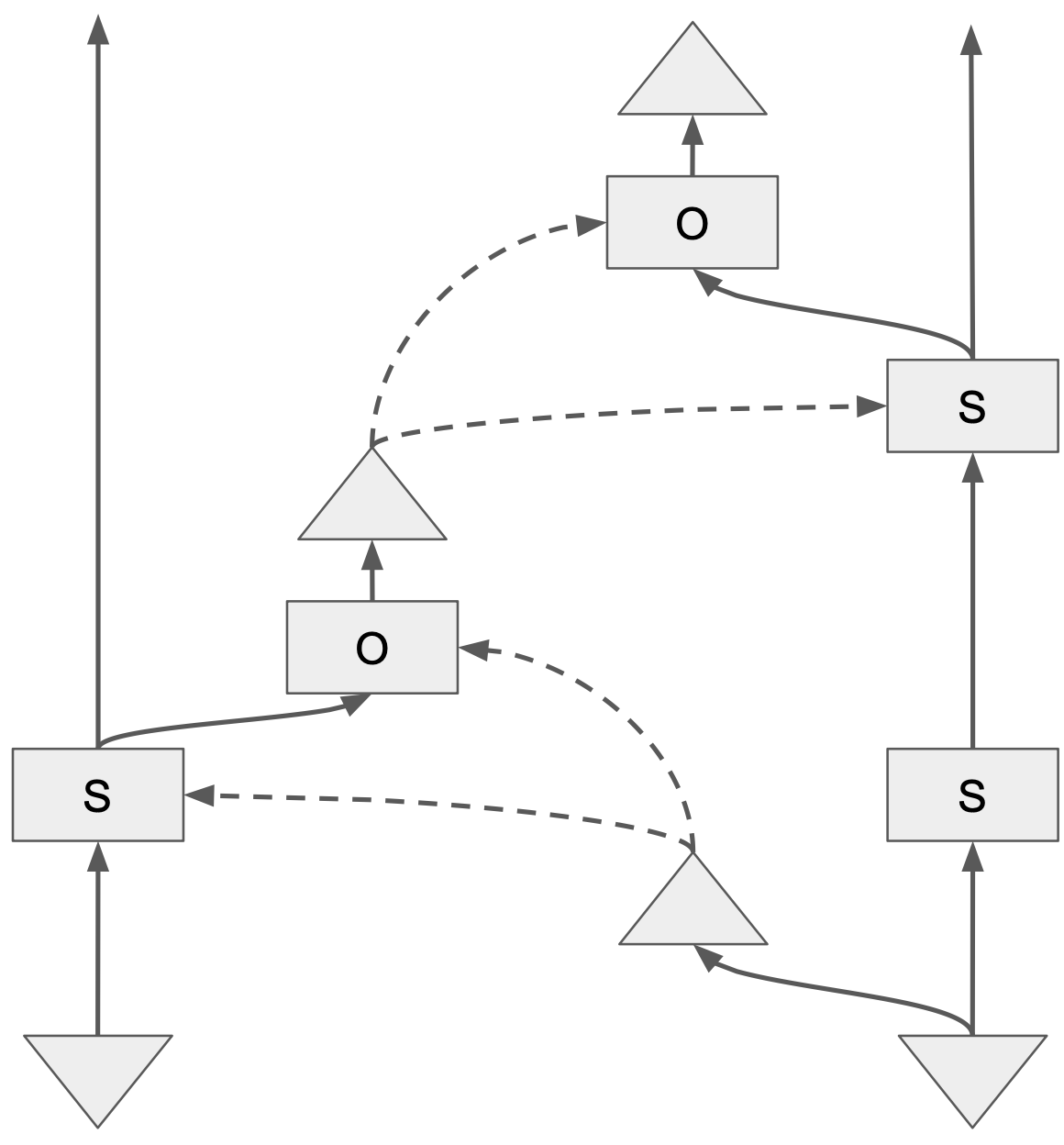}
\end{figure}

Here we see a process of repeated, reciprocal interactions. The stimulus from the prior diagram now produces some output from the LSP which then applies an $\{O,S\}$ pair to the RSP, which respectively both evolves the RSP's cogit vector and produces an output which again applies some general $\{S,O\}$ pair to the LSP's cogit vector. 

All $S$ and $O$ in this diagram, and indeed all processes in all diagrams within the psychodynamics process theory, are in fact operator-valued distributions (OVDs). While a full analysis of the algorithmic problem of inferring properties of these OVDs is beyond the scope of this paper, we intend to give such an analysis in a follow-up paper. In brief, algorithms from the field of quantum control theory can be applied. While traditionally they have been used to design optimal operators for accomplishing some desired transform to a quantum state, they can just as easily be used to take a specific state transformation and infer the structure of the operator which accomplished the transformation. By applying this over some set of sufficiently similar experiments or natural observations, one can approximate the OVD(s) corresponding to a given process. 

It is important to note that one can define cogits in terms of the observed outputs in response to various $O$ operators. Because of this, the $O$ are actually the starting point for describing psychodynamic processes. This is one of the most helpful features of the psychodynamic processes, because it grounds all systems in terms of measurable properties. That said, it is not possible to have various $O$ correspond to random unitary matrices, and some set $\{O_i\}_{i=1,...n}$ must be defined in a fashion which are with high probability internally consistent. For example, "Are you hungry?" and "Do you feel hungry?" cannot be defined such that they give opposite answers with high probability. As such, one would need to begin with some specific operator $O_0$ and then progressively add additional operators such that the probabilities match empirical values. However, any given $O_i$ need not be so simple as "Are you hungry?" As a large-scale example, a battery of numerous questions could be approximated as a single operator, though a more fine-grained analysis would reveal that the ordering of the questions would to at least some degree influence the outcomes.\cite{Wang_2013_quantum_question_order_model,Bruza_2015_quantum_cognition_review,Pothos_2022_quantum_cognition_review,Busemeyer_2023_quantum_cognition_models} However, if the battery of questions were universally given in a consistent order, then it would be perfectly valid to approximate the entire battery as a single operator when considering processes with time-steps of durations longer than the period over which such a battery of questions were given. This hearkens back to the earlier discussion of the temporally scale-free nature of the mathematical framework of holographic cognition, as discussed at greater length in Bagley \& Petritsch.\cite{bagley_petritsch_2024_holographic_cognition_original_cognitive_neurosecurity}

\section{W-L Language Games}

Armed with this process theory, we turn to central example which further illustrates its use. A very general problem would be the form of "language game" described by Wittgenstein.\cite{Wittgenstein1968Philosophical} Two parties each have a "private language" which both reflects, is the product of, and generates (at least some aspects of) their understandings of the world, internal lives, and experiences. There is a public language, which is the means by which the two parties communicate, and the goal of one or both parties is to gain insight into the private language of the other. In honor of Wittgenstein's use of a lion as an example of entities which he supposed would possess a private language which humans could never understand, we will term these two parties Lion and Wittgenstein, denoted L and W in the diagrams. 

W would like to infer information about both the cogit vector state as well as the OVD(s) corresponding to the evolution of L's cogit vector. The public language of the $O$ operators is a shared ground-truth, as is the basis which it fixes for both W's and L's respective cogit vectors. However, at initialization of the W-L game W has no information about L's cogit state $\mathcal{L}_0$. Some subset of W's cogit state $\mathcal{W}_0$ corresponds to their Bayesian priors for both $\mathcal{L}_0$ and the OVD(s) of the $S(\mathcal{L})_t$ which would be applied by the presentation by $W$ at time $t$ of measurement operators $O_i$ from some set of options $i=\{1,...,N\}$.

For the sake of simplicity, let us assume a case where W's initial hypothesis $\mathcal{W}_0$, a subset of W's cogit vector, is a completely uniform prior over all possible values of the cogit vector $\mathcal{L}_0$. Additionally, we assume that L either cannot deceive W, or is sufficiently disincentivized against lying that it chooses to provide truthful responses.\footnote{An analysis of cases with deception is beyond the scope of this paper, and will be assessed in future work.} For the sake of realism let us assume that only a subset of $\mathcal{L}_t$ can actually be read out by any given $O_i$, and that the set cogits $\mathcal{L}$ which could even possibly be measured by $\cup_i O_i$ is a strict subset of some larger cogit collection describing the whole of L's cognition. That is, $\mathcal{L}\subset L$. Let us additionally assume that all $O_i$ have the same upper bound on the Bures distance $B(\tilde{\mathcal{L}}_0,\mathcal{L}_0)$.\footnote{See Bagley \& Petritsch 2024 for a discussion of why the Bures distance is used in this context.\cite{bagley_petritsch_2024_holographic_cognition_original_cognitive_neurosecurity}} In this case, the choice of $O_i \rightarrow O_{t=1}$ can be random across the set of available $O$ while remaining optimal. For W, the hypothesis 

Because the basis of the cogit vectors for W and L are the same--fixed by their relationship with the public language of the $O_i$--the true objective of the inference problem is to reduce the entropy of the OVD over $S_L(O_i)_t$. W thereby reduces uncertainty with respect to their hypothesis of the private language of L.

There is thus an output sampled from the distribution $T_1 = O_i(\mathcal{L}_0)_{t=1}$, which applies a dynamics operator $S_{W,1}$ to $\mathcal{W}_1$ to produce $\mathcal{W}_2$. 

This leads us to a more formal mathematical description of the inference problem underlying W-L games.

\subsection{Formulation of W-L Games as Hierarchical Bayesian Inference}
Here we formalize the Wittgenstein Lion (W-L) Language game as an entropy reduction problem within the holographic cognition framework. We use hierarchical Bayesian modeling to show how Wittgenstein (W) infers the cognitive dynamics or private language of an individual L through the use of properties of a population. As W interacts with the L, we recursively update a joint posterior belief over both the L's hidden cognitive state and their unique OVD parameters. We define the objective function for an algorithm aimed at deriving an individual's OVD based on expected information gain (EIG), which allows us to actively select the next question that maximally reduces our uncertainty (entropy) about the individual L's specific OVDs, refining the population prior to characterize the individual's private language.

\subsubsection{Cognitive States and Dynamics}

The state space of the cognitive system of the L is modeled within a \(D\)-dimensional complex Hilbert space equipped with an inner product, denoted \(\mathcal{H_D} \cong \mathbb{C}^D\). Let \(\mathcal{D}_D\) denote the space of density matrices on \(\mathcal{H}_D\).

\begin{defn}[Cognitive State]
    The mental state of the L at time \(t\) is represented by a density matrix \(\rho_t \in \mathcal{D}_D\). This is a positive semi-definite operator on \(\Hc_D\) \((\rho_t \succeq 0)\) with unit trace \((\Tr(\rho_t)=1)\).
\end{defn}

This provides us a complete description of the L's cognitive state at a given time, handling both the definite beliefs and statistical uncertainty of mixed states.

The observer (W) interacts with L using a finite set of public language operators, \(\Oc = \{O_1, \dots, O_N\}\).

\begin{defn}[Public Language Operators]
    Each operator \(O_i \in \Oc\) is formalized as a Positive Operator-Valued Measure (POVM). A POVM is a set of elements \(\{E_{i,a}\}_a\), where \(a\) indexes the possible outcomes (Answers, A). These elements satisfy \(E_{i,a} \succeq 0\) and the completeness relation \(\sum_a E_{i,a} = \Id \).
\end{defn}

We model cognitive evolution, which is the change in mental state due to internal processes \(S\) or external stimuli, through the application of unitary operators. These operators belong to the Unitary Group \(U(D)\), which means that the evolution preserves the probabilistic structure, and the state remains normalized. It should be noted that for any given set of empirically measurable cogits there will always be unmeasured or unmeasurable cogits--at least in any real-world setting. This means that operators describing cognitive dynamics would not truly be unitary. However, one can add "junk" cogits to the state vector as an approximate way of in some sense absorbing the non-unitarity. This is a traditional approximation from quantum mechanics when dealing with a mixed state--i.e. one in which the qubits are entangled not only within the system but also with the environment surrounding the system. In holographic cognition, the analogy to the "environment" is all portions of an individual's cognition not captured by a given set of cogits. 

The probability of observing outcome \(a\) when applying a public operator \(O_i\) to state \(\rho_t\) is determined by the Born rule:

\begin{equation} 
    P(a | O_i, \rho_t) = \Tr(E_{i,a} \rho_t).
	\label{eq:born_rule}
\end{equation}


This is the probability that the L will give a specific answer when W asks a question, given L's current mental state.

\subsubsection{Operator-Valued Distributions }

Applying a stimulus \(O_i\) induces a dynamic evolution of the cognitive state, giving us \(\rho'\). We model this as a stochastic evolution, where we draw from a distribution of unitaries: 

\begin{equation}
	\rho' = U \rho U^\dagger
	\label{eq:unitary_evolution}
\end{equation}

where \(U\) is a unitary matrix drawn from the Unitary group \(U(D)\), and \(U^\dagger\) is its conjugate transpose. The specific transformation \(U\) is stochastic and governed by an OVD associated with \(O_i\). This describes how L's mental state changes. When L processes a stimulus, their internal state \(\rho\) is transformed into a new state \(\rho'\) by a cognitive process \(U\).


\begin{defn}[Operator-Valued Distribution]
    An OVD \(\Pc_i\) is a probability distribution over \(U(D)\), characterizing the stochastic cognitive dynamics induced by the operator \(O_i\).
\end{defn}

We assume the OVDs belong to a parameterized family of distributions \(\Pc_i(U | \theta_i)\). The parameters \(\theta_i\) belong to some parameter space \(\Omega_{\theta_i}\).
	
\begin{equation}
	U \sim \Pc_i(U | \theta_i).
    \label{eq:ovd_sampling}
\end{equation}


This is the specific cognitive process \(U\) that occurs when W asks question \(O_i\). It is drawn from a distribution \(\Pc_i\) that characterizes L's private language defined by \(\theta_i\).

The L's private language is characterized by the complete set of these parameters \(\bm{\Theta} = \{\theta_1, \dots, \theta_N\}\), which exist in the joint parameter space \(\Omega_{\Theta} = \Omega_{\theta_1} \times \dots \times \Omega_{\theta_N}\). The primary objective is the inference of \(\bm{\Theta}\).

\subsection{The Hierarchical Bayesian Framework}
We use a hierarchical Bayesian strategy to manage individual variability and the complexity of hidden state dynamics. We start with analysis on the population, which gives us a prior for analysis at the individual level.

\begin{enumerate}
	\item \textbf{Population Level:} W first establishes a population-level prior \(P_{pop}(\bm{\Theta})\) by analyzing interactions across a population of Ls. This estimates the general distribution of OVD parameters, marginalizing over individual variations and unknown initial states.
		
	\item \textbf{Individual Level:} W then engages with a specific individual. The population distribution serves as the initial prior belief: \(P(\bm{\Theta}|H_0) = P_{pop}(\bm{\Theta})\). Here, \(H_0\) represents the initial condition which is the state of knowledge before any interactions have occured with a specific individual L. We also assume a prior over the initial cognitive state \(P(\rho_0)\) (e.g., the maximally mixed state \(\Id/D\)).
\end{enumerate}

It is important to note that the maximally mixed state prior \(\Id/D\) is the prior over the initial cognitive state, rather than using information from the population level. The hierarchical model informs the dynamics \(\Theta\) which describe how the mind changes. It does not necessarily inform the initial state \(\rho_0\) of a specific individual at the start of the interaction. The cogit states of each individual are unknown variables.

The dialogue proceeds sequentially. Let \(H_t = \{(O_1, A_1), \dots, (O_t, A_t)\}\) denote the history of interactions up to time \(t\). At step \(t+1\):
	
\begin{enumerate}
    \item W selects \(O_{t+1} \in \Oc\).
	\item A hidden unitary \(U_{t+1} \sim \Pc_{O_{t+1}}(U | \theta_{O_{t+1}})\) is realized.
    \item L's state evolves: \(\rho_{t+1} = U_{t+1} \rho_t U_{t+1}^\dagger\).
	\item L generates a response \(A_{t+1}\) based on \(\rho_{t+1}\) via the Born rule (\Cref{eq:born_rule}).
\end{enumerate}
	
The inference requires the simultaneous estimation of the static parameters \(\bm{\Theta}\) and the dynamic hidden state \(\rho_t\). W must maintain and update a joint posterior belief distribution \(P(\bm{\Theta}, \rho_t | H_t)\). This distribution represents the updated belief about both the static parameters \(\Theta\) and the dynamic cognitive state \(\rho_t\), conditioned on the observed history \(H_t\). The state and parameters are strongly interdependent. The observations \(A_t\) depend directly on the current state \(\rho_t\) via the Born rule, and the evolution of  \(\rho_t\) to \(\rho_{t+1}\) depends directly on the parameters \(\Theta\) via the OVD. The joint distribution allows us to capture the correlations and shared uncertainty between these variables for more accurate inference. 

\subsubsection{Recursive Bayesian Update}

We derive the recursive update for the joint belief state when W applies \(O_{t+1}\) and observes \(A_{t+1}\). First, we calculate the predictive distribution \(P(\bm{\Theta}, \rho_{t+1} | O_{t+1}, H_t)\) using the Chapman-Kolmogorov equation, marginalizing over the previous state \(\rho_t\):

\begin{equation}
P(\bm{\Theta}, \rho_{t+1} | O_{t+1}, H_t) = \int_{\mathcal{D}_D} P(\rho_{t+1} | \bm{\Theta}, \rho_t, O_{t+1}) P(\bm{\Theta}, \rho_t | H_t) d\rho_t.
\label{eq:prediction_step}
\end{equation}


Before W hears the answer to the next question \(O_{t+1}\), W predicts the new mental state \(\rho_{t+1}\) and the private language parameters \(\Theta\). This involves considering all possible current mental states \(\rho_t\) and how they would evolve according to the hypothesized dynamics. 

The state transition probability \(P(\rho_{t+1} | \bm{\Theta}, \rho_t, O_{t+1})\) encapsulates the dynamics of the stochastic evolution. It is calculated by integrating over all possible unitary realizations \(S \in U(D)\):

\begin{equation} 
    P(\rho_{t+1} | \bm{\Theta}, \rho_t, O_{t+1}) = \int_{U(D)} \delta(\rho_{t+1} - U\rho_t U^\dagger) \Pc_{O_{t+1}}(U | \theta_{O_{t+1}}) dU,
	\label{eq:state_transition}
\end{equation}

	
where \(\delta(\cdot)\) is the Dirac delta function generalized for the space of density matrices. For reference, The standard Dirac delta \(\delta(x-a)\) is a generalized function that is zero everywhere except at \(x = a\), and its integral over the domain is 1. In the above equation, the Dirac delta is generalized to the space of density matrices: 

\[\delta(\rho_{t+1} - U\rho_t U^\dagger)\] 

This generalized function enforces the constraint of unitary evolution. When integrated, it yields a non-zero value only if \(\rho_{t+1}\) is exactly equal to the result of applying the specific unitary \(U\) to \(\rho_t\). It acts as a selector, enforcing the deterministic relationship between \(\rho_t\), \(U\), and \(\rho_{t+1}\) for a specific realization of \(U\).

The next step is the correction and measurement update. Upon observing \(A_{t+1}\), the belief is updated using Bayes' theorem. The new history is \(H_{t+1} = (H_t, O_{t+1}, A_{t+1})\).
	
\begin{equation} 
    P(\bm{\Theta}, \rho_{t+1} | H_{t+1}) = \frac{P(A_{t+1} | \bm{\Theta}, \rho_{t+1}, O_{t+1}, H_t) P(\bm{\Theta}, \rho_{t+1} | O_{t+1}, H_t)}{P(A_{t+1} | O_{t+1}, H_t)}.
	\label{eq:correction_step}
\end{equation}


Here, W observes L's answer \(A_{t+1}\) and updates their belief. Hypotheses that predicted the observed answer well become more probable, while those that did not become less probable.
	
The likelihood of the observation exhibits the Markov property and is given by the Born rule:

\begin{equation} 
	P(A_{t+1} | \bm{\Theta}, \rho_{t+1}, O_{t+1}, H_t) = P(A_{t+1} | O_{t+1}, \rho_{t+1}) = \Tr(E_{O_{t+1}, A_{t+1}} \rho_{t+1}).
	\label{eq:likelihood}
\end{equation}

It is important to clarify what the Markov property means here. This assumption states that the future is conditionally independent of the past, given its current state. This appears in \Cref{eq:prediction_step}, \Cref{eq:state_transition}, and \Cref{eq:likelihood}. It means that if we know the current cognitive state \(\rho_{t+1}\), the history \(H_t\) provides no additional information about the observation \(A_{t+1}\). The density matrix \(\rho_t\) is defined as the complete representation of the cognitive state, and in the holographic cognition framework, \(\rho_t\) encapsulates all relevant information, including memories, beliefs, traumas, and subconscious elements, as described in section 2.2 of this paper. This assumption does not say that the L forgets its past, but rather the L's entire relevant past is summarized within its current mental state \(\rho_t\).
	
The denominator is the marginal likelihood (evidence) of the observation:

\begin{equation}
    P(A_{t+1} | O_{t+1}, H_t) = \int_{\Omega_{\Theta}}\int_{\mathcal{D}_D} P(A_{t+1} | O_{t+1}, \rho_{t+1}) P(\bm{\Theta}, \rho_{t+1} | O_{t+1}, H_t) d\rho_{t+1} d\bm{\Theta}.
    \label{eq:marginal_likelihood}
\end{equation}


This is the overall probability of observing the answer \(A_{t+1}\), averaged across all of W's hypotheses about L's mental state and private language.

\subsection{Optimal Experiment Design for Entropy Reduction}

W aims to select the sequence of operators \(\{O_t\}\) that maximizes the efficiency of learning \(\bm{\Theta}\). This is formalized as a Bayesian optimal experiment design (OED) problem focused on uncertainty reduction.

The primary objective is the inference of the OVD parameters. We define the marginal parameter posterior \(P(\bm{\Theta}|H_t)\) by integrating out the cognitive state:
	
\begin{equation}
    P(\bm{\Theta}|H_t) = \int_{\mathcal{D}_D} P(\bm{\Theta}, \rho_t | H_t) d\rho_t.
    \label{eq:marginal_posterior}
\end{equation}


To focus on the private language \(\bm{\Theta}\) rather than the momentary mental state \(\rho_t\), W calculates the belief about \(\bm{\Theta}\) by averaging over all possibilities for the hidden mental state \(\rho_t\).

Uncertainty is quantified by the differential entropy \(H\) of this marginal posterior:

\begin{equation}
    H(\bm{\Theta}_t) \equiv H(\bm{\Theta} | H_t) = - \int_{\Omega_{\Theta}} P(\bm{\Theta} | H_t) \log P(\bm{\Theta} | H_t) d\bm{\Theta}.
    \label{eq:entropy}
\end{equation}


W employs an active learning strategy by selecting the operator \(O_{t+1}\) that maximizes the expected information gain (EIG), defined as the expected reduction in entropy:

\begin{equation} 
    EIG(O_{t+1} | H_t) = H(\bm{\Theta}_t) - \E_{A_{t+1}|O_{t+1}, H_t}[H(\bm{\Theta}_{t+1})].
	\label{eq:eig_definition}
\end{equation}


This calculates how much W expects their uncertainty to decrease, on average, by asking a specific question \(O_{t+1}\).
	
The expectation is taken over the potential outcomes \(A_{t+1}\), weighted by their predictive probability (\Cref{eq:marginal_likelihood}). The optimal action policy (in a greedy sense) is:

\begin{equation} 
	O_{t+1}^* = \arg\max_{O_i \in \Oc} EIG(O_i | H_t).
	\label{eq:optimal_action}
\end{equation}

This objective function selects the operator \(O_{t+1}\) that is expected to yield the most information about the parameters \(\Theta\), thereby minimizing the entropy of the OVDs. We employ a greedy approach here due to computational necessity. The truly optimal policy would require anticipating all possible future sequences of questions and answers, which would not be feasible for a high-dimensional continuous state.  

The EIG is equivalent to the Mutual Information \(I(\bm{\Theta}; A_{t+1} | O_{t+1}, H_t)\). This can be reformulated in terms of the Kullback-Leibler (KL) divergence:
	
\begin{equation}
    EIG(O_{t+1} | H_t) = \int_{\Omega_{\Theta}} P(\bm{\Theta}|H_t) \left[ \sum_{A_{t+1}} P(A_{t+1}|\bm{\Theta}, O_{t+1}, H_t) \log \frac{P(A_{t+1}|\bm{\Theta}, O_{t+1}, H_t)}{P(A_{t+1}|O_{t+1}, H_t)} \right] d\bm{\Theta}.
    \label{eq:eig_expanded}
\end{equation}

This leads to the interpretation of EIG as the expected KL divergence between the likelihood predicted by a specific model realization \(\bm{\Theta}\) and the marginalized likelihood:
	
\begin{equation} 
	EIG(O_{t+1} | H_t) = \E_{\bm{\Theta}|H_t} \left[ \KL\left( P(A_{t+1}|\bm{\Theta}, O_{t+1}, H_t) \ || \ P(A_{t+1}|O_{t+1}, H_t) \right) \right].
	\label{eq:eig_kl}
\end{equation}


This formulation shows that expected information gain is maximized when the predictions of different plausible models \(\bm{\Theta}\) are highly varied compared to the average prediction. W seeks the question that best discriminates between competing hypotheses. 

This criterion favors operators for which the predictions of different plausible models \(\bm{\Theta}\) are most divergent, thereby maximizing the ability to discriminate between hypotheses.

Evaluating the EIG requires computing the model-specific likelihood terms \(P(A_{t+1}|\bm{\Theta}, O_{t+1}, H_t)\). This necessitates marginalization over the hidden cognitive states:

\begin{equation}
    P(A_{t+1}|\bm{\Theta}, O_{t+1}, H_t) = \int_{\mathcal{D}_D} P(A_{t+1}|O_{t+1}, \rho_{t+1}) P(\rho_{t+1}|\bm{\Theta}, O_{t+1}, H_t) d\rho_{t+1}.
    \label{eq:model_likelihood}
\end{equation}

	
The term \(P(\rho_{t+1}|\bm{\Theta}, O_{t+1}, H_t)\) is the state prediction conditioned on a specific model \(\bm{\Theta}\):

\begin{equation}
    P(\rho_{t+1}|\bm{\Theta}, O_{t+1}, H_t) = \int_{\mathcal{D}_D} P(\rho_{t+1} | \bm{\Theta}, \rho_t, O_{t+1}) P(\rho_t|\bm{\Theta}, H_t) d\rho_t.
    \label{eq:conditioned_prediction}
\end{equation}

This calculates the probability distribution over the next mental state \(\rho_{t+1}\), assuming a specific private language \(\bm{\Theta}\) is true. It propagates the uncertainty about the current state \(\rho_t\) given \(\bm{\Theta}\) through the dynamics defined by \(\bm{\Theta}\).

\section{Connections to AI Safety}\label{section_connections_AI_safety}

In this section we lay out a series of potential applications for the methods above to problems in AI safety. Some are immediately feasible, others highly plausible, and in the case of the final example more speculative.

\subsection{Analysis of AI-driven Attacks on Cognition}

Suppose W were instead a misaligned or maliciously designed agent, and its goal was not only to infer the structure of L's private language but then also use it to select an optimal sequence of $O_t$ in order to selectively steer L's cogit vector towards some goal state which serves the objectives of the agent. It should be stressed that this is not some fantastical notion of brainwashing or indoctrination. Examples of such manipulation could extend to extremes of that sort, but are much more likely to be along the lines of more subtle effects like political manipulation, instilling desire for products, alteration of preferences, or gradual shifts in beliefs. All of these can be seen around us constantly in the forms of advertisements, media programs, and more. All such inputs to our cognition are public language operators,, so an AI agent would merely need to learn how to engineer content (e.g. via generative AI models) to optimally steer a target individual's or population's cognition by leveraging its model(s) of their private language(s).

Out of a desire to avoid introducing any nomenclature which might unjustly associate Ludwig Wittgenstein with misaligned and malicious agents, we will instead select a more fitting acronym. Given a radically different Ludwig--Ludwig Mueller--played a central role in efforts to rewrite German religious beliefs into Nazi propaganda, his surname appears fitting. 

In such an M-L game, the W-L game is extended with an additional quantum optimal control problem. Via inference over L's private language and cogits using the joint belief state analysis described above, M then attempts to leverage existing O (in the simplified discrete case above) or design novel O by solving an optimal quantum control problem over the Unitary Group U(d) and translating them into whatever form of stimuli are available as vectors for applying the O designed in this fashion. This phrasing is quite general for a reason. In the context of an agent which interacts via language, speech or text would be the stimuli. If an agent producing imagery or other sounds, the corresponding senses. Yet this generalizes further, as in the contexts of brain-computer interfaces and other neurotechnology, the stimuli can be even more varied and far more direct. In each case, the general mathematical structure should be unchanged from what we describe here, and indeed that generality is a deliberate part of the approach we have described. 

Let us now take this a step further in two directions. 
 
First, suppose a group of M, or a single M agent acting via multiple channels, aimed to collectively steer an individual's cogit vector towards some goal state. If framed in terms of stimuli transmitted to L, the single-M case may be easier to detect given it will need to be a single, stronger signal. By contrast, numerous less-impactful signals may be easier to disguise as noise by hiding them amidst other channels and signals--which may themselves even be transmitting information which has been inferred by M to optimize the tradeoff between obscuring the interference and hampering that same interference. 

One could mitigate this difficulty by instead attempting to monitor the cogit dynamics of L. Yet this presents its own quandry, as in the absence of a method for doing so \textit{without} violating L's privacy, one opens the door to further abuse--if not simply making the objectives of M agents easier to achieve. 

Second, suppose an M agent or group of M were to target the cognition of a group of L. Given a sufficiently accurate model of the dynamics by which interactions between Ls shape the cognition of others within the group, M gain(s) the option of further obscuring its influence by leveraging network effects within the group's social dynamics.\footnote{See\cref{appendix_group_cogit_bipartite} for a note on an open question regarding a relevant property of holographic cognition in social contexts.}

Full analysis of these problems is beyond the scope of this paper, but we introduce the notions here to help clarify the relevance of this paper to multiple problem spaces.

\subsection{Application to AI Alignment}

A central problem in AI alignment is the optimization of some function reflecting the discrepancy between human values and behavioral properties of AI systems. While this problem is immensely complex and multifaceted, one aspect of this complexity comes from the supposed "irrationality" of human beliefs, preferences, values, decisions, and so on. However, these properties are precisely those which have been empirically shown to be effectively modeled by cogits, and they are irrational only through the lens of classical probability. While by no means a panacea or magic bullet, the use of holographic cognition--and particularly the W-L game--provides a way to infer a mathematical model of cogits and psychodynamics at scale. Most importantly, such a model reflects the complex projective probability underlying human cognition, in contrast with the approach of modeling cognition via classical probability which has historically been the norm in AI alignment.

\subsection{Representation Agents}
There has been growing interest in the concept of AI agents capable of serving as representatives of human users in settings such as planning, negotiating sales or contracts, information exchange, and more. Some of these tasks can rely on purely logical, rule-based protocols, making them amenable to existing methods (so long as "bullshit" by the agents in question is sufficiently minimized and/or mitigated).\footnote{We use the term "bullshit" (as described in the philosophy literature after being introduced by Frankfurt to mean speech without any concern for truth value) rather than the term "hallucination" which is more popular within the AI literature.\cite{frankfurt2009bullshit} This is because "hallucination" assigns to AI anthropomorphic properties which are not based in evidence, inaccurately conflating a pathological internal process with expressed assertions about reality, and attempts to falsely transfer responsibility for malfunctions and failures of products from their creators to the products themselves. This obfuscation is highly counterproductive, as it distracts from the underlying source of such failure modalities being the design and engineering of models themselves.\cite{Fisher2024_bullshit,Birhane2024_bullshit,Hicks_2024_ChatGPT_bullshit,Tigard2025_bullshit,Sparrow2025_bullshit,Coeckelbergh2025_bullshit}}

By contrast, tasks like contract negotiation are more amorphous, and effectively carrying them out is more contingent on an agent having an effective model of the cogit state(s) and private language(s) of the human(s) whom it aims to represent. Inference of such a mathematical model via a W-L game can reflect the empirically correct probability structure, though it remains to be seen what applied methods would be required in order to accomplish this with sufficiently high accuracy and comprehensiveness. It will also be important to determine how much an individual's private language changes over time, and at what rate, in order to estimate how effective a model inferred via W-L game can be. Additionally, it will likely be important to time-stamp any such private language model, so that both the agent using the model and other agents with which it might interact can all have information as to the recency--and thus viability--of the private language model in question.

If such applied problems can be effectively resolved, we anticipate two branches for using W-L games to train agents to effectively represent individuals.

\subsubsection{Cognitive Dossier}

In the first, one imagines an agent using the private language model obtained via W-L game as a guidebook of sorts. Much as one might use a dossier on an individual to gain insight into their cogits, and could leverage this to represent that individual, an AI agent could reference this model to estimate how effectively different courses of action align with the cogits of the individual whom the agent is tasked to represent. 

Another effective metaphor is that of a will--a document meant to reflect and describe the intent and preferences of an individual in their absence. 

\subsubsection{Gogol}

Mathematically similar yet functionally quite different is the notion of a gogol. A concept introduced by the physicist and author Hannu Ranjaniemi, a gogol is an approximate duplicate of a human mind. Such entities operate semi-autonomously or fully so, carrying out a variety of tasks.\cite{Rajaniemi_2010_the_quantum_thief,Rajaniemi_2012_the_fractal_prince,Rajaniemi_2014_the_causal_angel} 

It is interesting that Ranjaniemi describes gogols as existing on quantum computers, as there does not appear to be any indication he was aware of the use of complex projective probability to model cognition, and the mathematics necessary for constructing a gogol were completely absent at the time--indeed being first derived in Bagley \& Petritsch and this paper.\cite{bagley_petritsch_2024_holographic_cognition_original_cognitive_neurosecurity}

While most of the applications of gogols proposed by Ranjaniemi are highly speculative, there is at least one very concrete class of applications which are much more imminently feasible--assuming quantum computers with sufficiently high logical qubit counts. Though quantum cognition--and thus holographic cognition--shares the same form of probability as quantum mechanics, because there is no connection to quantum physics itself there is no analog of the fundamentally physical properties which enable certain quantum protocols. For such cases, creating a gogol and translating it to a quantum computer could enable necessarily-quantum algorithms and protocols to operate on models of human cognition. Examples could include quantum teleportation, quantum games, or secure multi-party computations such as in a quantum Trusted Execution Environment. If the error tolerance of a desired application and the minimum required cogit count are sufficiently non-demanding, it may be possible to implement such methods in the very near future. 

\section{Discussion}

In this paper we have provided a mathematically precise formulation of psychodynamics and general cognitive processes via a modification of the Symmetric Monoidal Category used to describe quantum probability. Thanks to the advantageous properties of the diagrammatic form of category theory and especially process theories, this formulation and its notation will hopefully be easily interpretable to experts from a range of fields. The fact it is mathematically precise and rigorous despite this approachability is the greatest advantage of the psychodynamic process theory we present here, as such mathematical rigor will be essentual for many of the most urgent and critical applications that we forsee.

Second, we formulated the W-L game, a model of the problem of inferring another party's cognitive states and dynamics which is adapted from the language games of Ordinary Language Philosophy. Third, we analyzed the Bayesian optimization problem which reflects the core of a W-L game from a quantitative perspective. While a full discussion of algorithmically solving this problem are beyond the scope of this paper, we intend to explore them in follow-up work. This connects to the potential applications discussed in Section \cref{section_connections_AI_safety}, and given the correct experimental environment, along with algorithms which are sufficiently efficient for the scale of data to be studied, this provides a protocol for solving inference problems with the potential for significant impact on multiple problems in AI safety. Further application to clinical psychology and psychiatry is also highly plausible, and we are in the process of developing these connections with collaborators from psychiatry. 

Having expanded upon the framework of holographic cognition, here we have presented a more concrete exploration of its properties and use across multiple domains relevant to human health and AI safety, areas where it holds great potential for application.

\newpage

\noindent\textbf{Acknowledgements}\\
We would like to thank Nicholas Godfrey, PhD, for helpful discussions in the planning and ideation stages of this paper. We would also like to thank Dany Lamothe, MD, and Daniel Fishman, MD for discussions of the application of holographic models of cognition to psychodynamic psychotherapy, as well as feedback on the relationship between psychotherapy and W-L games. Finally, we thank Claudia Petritsch, PhD, for her mentorship and support of the Mathematical Medicine Group.\\

\noindent\textbf{Author Contributions}
B.A. Bagley: Project ideation, development of the process theory and notation, analysis of the Bayesian optimization problem, writing of the paper; N. Khoshnan: Analysis of the Bayesian optimization problem, writing of the paper.

\newpage
\printbibliography

\newpage

\appendix

\section{Entanglement Structures in Individual and Group Cogit Vectors}\label{appendix_group_cogit_bipartite}

There is an interesting question which arises from the structure of holographic cognition, and which is made particularly clear in the category-theoretic formulation of the same. As discussed in \cite{bagley_petritsch_2024_holographic_cognition_original_cognitive_neurosecurity}, holographic cognition is scale-free, or perhaps more precisely scale-flexible, with respect to both time and the space of possible cogit hypervectors. One could use operators to reflect changes across any time range of interest, and can likewise construct a cogit hypervector so that it is a concatenation of the cogits from multiple individuals, scaling to however many individuals one might wish to study.

However, is such a concatenation actually the correct description of a group's collective state? Phrased differently, we know that the entanglement structure for a single individual scales with dimension in such a way that there is a rapid decline in the probability of being able to partition the cogits within their hypervector into fully non-entangled subsets. At a group level, the null hypothesis is that such partitionings of a group cogit hypervector would be along the same lines as reversing a simple concatenation. As such, interactions between the cogit hypervectors of different individuals would in general reflect graphs over which classical information processes are occurring. 

But is that true in the general case? An entanglement structure represents some collection of properties in the informational content and structure of an individual's nervous system, and it would be perfectly reasonable--indeed more plausible--for there to be no useful notion of entanglement structures bridging different individuals. Yet there are known phenomena which suggest it may be possible for the same mathematical structure to persist between individuals. It has been shown in the experimental neurosciene literature that certain patterns of neural activity become more similar in individuals with certain sorts of shared traits as those individuals interact. IF this or other phenomena result in a group cogit hypervector no longer being partitionable into hypervectors corresponding to the individuals, there would be an entanglement structure representing the group. 

There is no reason a priori to expect that this latter option is the case. While heuristic analogies of societies as intelligent superorganisms have existed for some time, there is no reason to think they would reflect underlying mathematical and information properties in a way directly equivalent to a rescaling of an individual's mind. However, if group cogit hypervectors are in general not partitionable, this would demonstrate such heuristic analogies have instead stumbled upon something quite substantive from a computational and phenomenological perspective. In mentioning this, we are not hypothesizing that group cogit hypervectors cannot be partitioned. Rather, that this is a property which should be quite amenable to experimental study, and could have significant and fascinating implications for quantifying phenomena at the interfaces of psychology and sociology.

\end{document}